\documentclass[10pt,a4paper,english]{article}

\usepackage[left=2.3cm, top=2.5cm, bottom=2.5cm, right=2.3cm]{geometry}
\usepackage{amsfonts}
\usepackage{amsmath}
\usepackage{amssymb,bm}
\usepackage{babel}
\usepackage[utf8]{inputenc}

\providecommand{\U}[1]{\protect\rule{.1in}{.1in}}
\numberwithin{equation}{section}
\providecommand{\U}[1]{\protect\rule{.1in}{.1in}}
\newtheorem{theorem}{Theorem}

\newtheorem{example}[theorem]{Example}

\newtheorem{lemma}[theorem]{Lemma}

\newtheorem{remark}[theorem]{Remark}

\newenvironment{proof}[1][Proof]{\noindent\textbf{#1.} }{\ \rule{0.5em}{0.5em}}

\begin{document}

\title{\sc St\"{a}ckel representations of stationary KdV systems}

\author{Maciej B\l aszak$^\dagger$, B\l a\.zej M. Szablikowski$^\dagger$
and Krzysztof Marciniak$^\ddagger$\\[1em]
\small $^\dagger$Department of Mathematical Physics and Computer Modelling\\
\small Faculty of Physics, Adam Mickiewicz University\\
\small Uniwersytetu Pozna\'nskiego 2, 61-614 Pozna\'{n}, Poland\\
\small\tt blaszakm@amu.edu.pl, bszablik@amu.edu.pl\\[0.75em]
\small $^\ddagger$Department of Science and Technology\\
\small Link\"{o}ping University, Campus Norrk\"{o}ping\\
\small 601 74 Norrk\"{o}ping, Sweden\\
\small\tt krzma@itn.liu.se}

\maketitle

\begin{abstract}
In this article we study St\"{a}ckel representations of stationary
KdV systems. Using Lax formalism we prove that these systems have
two different representations as separable St\"{a}ckel systems of
Benenti type, related with different foliations of the stationary
manifold. We do it by constructing an explicit transformation
between the jet coordinates of stationary KdV systems and
separation variables of the corresponding Benenti systems for
arbitrary number of degrees of freedom. Moreover, on the stationary
manifold, we present the explicit form of Miura map between both
representations of stationary KdV systems, which also yields their
bi-Hamiltonian formulation.
\end{abstract}

\section{Introduction}

It is well known that various reductions of soliton hierarchies lead to
Liouville integrable finite-dimensional systems. Stationary flows, restricted flows, Lax
constrained flows are examples of such reductions (see survey \cite{blaszak1998} and the
literature therein). The KdV hierarchy is by far the most studied of soliton hierarchies,
also from the point of view of its reductions. Theory of its stationary flows
was studied since the early 70's. Its finite gap solutions were found by
Dubrovin and Novikov \cite{DN,D1,D2} and its Riemann theta function
representation was presented by Its and Matveev \cite{I1,I2} (see the
comprehensive survey \cite{GE} and the literature therein).
Bogoyavlenskii and Novikov proved \cite{novik76} that these flows have the structure of
finite-dimensional Hamiltonian systems, a result generalized to stationary flows of
other evolution equations by Mokhov in \cite{mokhov}. Al'Ber \cite{alber81} constructed
canonical coordinates for the stationary KdV flows based on the algebraic recursion for conserved 1-forms (co-symmetries)
established in \cite{alber79} (or, equivalently, on the integro-differential recursion
for the KdV hierarchy \cite{magri}). Consequently, the bi-Hamiltonian
formulation for the KdV stationary flows was presented by means of the
degenerate Poisson tensors \cite{ant87} and as result Liouville
integrability of the stationary KdV flows was fully proved. It was also
observed that, in fact, in the case of the KdV hierarchy there are two
Hamiltonian finite-dimensional representations of the stationary flows,
connected by the Miura map \cite{ant92,tondo,rauch96}. Further, the
bi-Hamiltonian structure of the stationary KdV flows and their separability
was comprehensively studied in \cite{magri00}.

In this article we revisit these ideas in a novel, systematic way. We prove that each stationary KdV
system (by which we mean a stationary flow of KdV hierarchy together with all lower flows)
has two different St\"{a}ckel representations from the Benenti
class \cite{ben1,ben2,blaszak2006}. We prove this by stitching together Lax
representations of (an integrated form of) a given stationary KdV system and the corresponding St\"{a}ckel
separable system, which allows us to construct the explicit transformation between
jet coordinates of KdV stationary system and separation variables (via Vi\`{e}te's coordinates) of the corresponding St\"{a}ckel system.
We also present an explicit formula for a Miura map between both St\"{a}ckel representations of the same KdV stationary system
which leads to its bi-Hamiltonian representation.

Let us also mention that the inverse construction is also possible.
Starting from a carefully chosen family of St\"{a}ckel systems one can
reconstruct the related hierarchies of stationary systems and hence
reconstruct the whole KdV hierarchy. This idea was explored for the first
time in \cite{blaszak2006,blaszak2008}.

Denote the infinite KdV hierarchy by $u_{t_{k}}=\mathcal{K}_{k}(u)$, $k=1,2,\ldots$. The main result of this article is contained in the following theorem.

\begin{theorem}
\label{thm1} The $n$-th stationary KdV system, which consists of first $n$
flows from the KdV hierarchy~\eqref{h0} and the $(n+1)$-st stationary flow of KdV, i.e.
\begin{equation}
u_{t_{1}}=\mathcal{K}_{1},\qquad u_{t_{2}}=\mathcal{K}_{2},\qquad\ldots,%
\qquad u_{t_{n}}=\mathcal{K}_{n},\qquad\mathcal{K}_{n+1}=0,  \label{A123}
\end{equation}
can be integrated once either to the form \eqref{L17a} or to the form \eqref{L18a}. Each form is equivalent, through the appropriately defined map \eqref{C8}, to a St\"{a}ckel system from Benenti class. In the first representation the corresponding St\"{a}ckel system is defined by the spectral (separation) curve
\begin{equation*}
\lambda^{2n+1}+c\lambda^{n}+\sum_{k=1}^{n}H_{k}\lambda^{n-k}=\mu^{2},
\end{equation*}
and in the second representation by the spectral curve
\begin{equation*}
\lambda^{2n}+\bar{c}\lambda^{-1}+\sum_{k=1}^{n}\bar{H}_{k}\lambda^{n-k}=\lambda
\mu^{2},
\end{equation*}
where $c$ and $\bar{c}$ are respective integration constants of stationary flow $K_{n+1}=0$. In both St\"{a}ckel representations the evolution equations generated by Hamiltonians
$(H_{1},\ldots,H_{n})$ and $(\bar{H}_{1},\ldots,\bar{H}_{n})$ are
mapped by the corresponding map \eqref{C8} to the evolution equations of integrated stationary systems \eqref{L17a} and \eqref{L18a}, respectively. Moreover, the two St\"{a}ckel representations, \eqref{L17a} and \eqref{L18a} are related by the finite-dimensional Miura map \eqref{M1} on the extended $(2n+1)$-dimensional phase space on which the stationary system \eqref{A123} is defined.
\end{theorem}

Further, in the article we carefully explain all the ingredients
of this theorem, including the formulas this theorem refers to.

Recently, the (isospectral) Lax representations for the whole
Benenti class of St\"{a}ckel systems were constructed
\cite{blaszak2019}. These matrix Lax equations belong to the class
of the so-called Mumford systems that are associated with the
separable systems with separation curves of hyperelliptic type
\cite{mum,van}. Using the Lax formalism developed in
\cite{blaszak2019} we are able in this article to construct, in an
explicit form and for arbitrary number of degrees of freedom,
transformations between jet coordinates of a given KdV stationary
system and separation variables of the associated St\"{a}ckel
systems, which is a new result. Let us emphasize that comparing to
the results for instance from \cite{alber81,tondo,magri00}, we not
only consider here the second representation of stationary KdV
systems, but also show the one-to-one equivalence between
evolution equations from the stationary systems and the
Hamiltonian evolution equations from the corresponding St\"{a}ckel
systems.

This article is organized as follows. In Section \ref{2} we remind
some important facts about the KdV hierarchy, its Lax- and
zero-curvature representations. Next, we define the notion of KdV
stationary systems and obtain their two representations together
with their respective Lax equations. In Section \ref{3} we present
basic facts about a particular class of St\"{a}ckel systems of
Benenti type as well as information about its Lax representation.
Finally, in Section~\ref{4}, we prove Theorem \ref{thm1}
formulated above and use the Miura map \eqref{M1} to construct a
bi-hamiltonian formulation for the stationary KdV system.

\section{KdV hierarchy \label{2}}

\subsection{Bi-Hamiltonian structure}

Let us collect some, important for further considerations, facts about the
KdV hierarchy. The KdV equation
\begin{equation*}
u_{t}=\frac{1}{4}u_{xxx}+\frac{3}{2}u u_{x}
\end{equation*}
is a member of the bi-Hamiltonian chain of nonlinear PDE's%
\begin{equation}
u_{t_{n}}\equiv\mathcal{K}_{n}=\pi_{0}d\mathcal{H}_{n}=\pi_{1}d\mathcal{H}%
_{n-1},\qquad n=1,2,...  \label{h0}
\end{equation}
where the two Poisson operators are
\begin{equation*}
\pi_{0}=\partial_{x},\qquad\pi_{1}=\frac{1}{4}\partial_{x}^{3}+\frac{1}{2}%
u\partial_{x}+\frac{1}{2}\partial_{x}u.
\end{equation*}
The hierarchy \eqref{h0} can be generated by the recursion operator and its
adjoint
\begin{equation*}
N\equiv\pi_{1}\pi_{0}^{-1}=\frac{1}{4}\partial_{x}^{2}+u+\frac{1}{2}%
u_{x}\partial_{x}^{-1},\qquad N^{\dagger}=\frac{1}{4}\partial_{x}^{2}+u-%
\frac{1}{2}\partial_{x}^{-1}u_{x},
\end{equation*}
in the sense that
\begin{equation}
\mathcal{K}_{n+1}=N^{n}\mathcal{K}_{1},\qquad\gamma_{n}=d\mathcal{H}_{n}=%
\bigl ( N^{\dagger}\bigr) ^{n}\gamma_{0},\qquad n=1,2,\ldots\,.  \label{h2}
\end{equation}
In particular, we find that that the first vector fields (symmetries) $\mathcal{K}_{n}$ are:
\begin{equation*}
\begin{split}
\mathcal{K}_{1} & =u_{x}, \\
\mathcal{K}_{2} & =\frac{1}{4}u_{xxx}+\frac{3}{2}u u_{x}, \\
\mathcal{K}_{3} & =\frac{1}{16}u_{5x}+\frac{5}{8}u u_{3x}+\frac{5}{4}%
u_{x}u_{xx}+\frac{15}{8}u^{2}u_{x}, \\
\mathcal{K}_{4} & =\frac{1}{64}u_{7x}+\frac{7}{32}u u_{5x}+\frac{21}{32}%
u_{x}u_{4x}+\frac{35}{32}u_{xx}u_{3x}+\frac{35}{32}u_{x}^{3}+\frac{35}{8}u
u_{x}u_{xx}+\frac{35}{32}u^{2}u_{3x}+\frac{35}{16}u^{3}u_{x}, \\
& \vdots
\end{split}%
\end{equation*}
the first conserved one-forms (co-symmetries) $\gamma_{n}$ are
\begin{equation*}
\begin{split}
\gamma_{0} & =2, \\
\gamma_{1} & =u, \\
\gamma_{2} & =\frac{1}{4}u_{xx}+\frac{3}{4}u^{2}, \\
\gamma_{3} & =\frac{1}{16}u_{4x}+\frac{5}{8}u u_{xx}+\frac{5}{16}u_{x}^{2}+%
\frac{5}{8}u^{3}, \\
\gamma_{4} & =\frac{1}{64}u_{6x}+\frac{7}{32}u u_{4x}+\frac{7}{16}%
u_{x}u_{3x}+\frac{21}{64}u_{xx}^{2}+\frac{35}{32}u^{2}u_{xx}+\frac{35}{32}%
uu_{x}^{2}+\frac{35}{64}u^{4}, \\
& \vdots
\end{split}%
\end{equation*}
while the first Hamiltonian densities $\mathcal{H}_{n}$ of conserved functionals are
\begin{equation*}
\begin{split}
\mathcal{H}_{0} & =2u, \\
\mathcal{H}_{1} & =\frac{1}{2}u^{2}, \\
\mathcal{H}_{2} & =-\frac{1}{8}u_{x}^{2}+\frac{1}{4}u^{3}, \\
\mathcal{H}_{3} & =\frac{1}{32}u_{xx}^{2}+\frac{5}{32}u^{2}u_{xx}+\frac {5}{%
32}u^{4}, \\
\mathcal{H}_{4} & =-\frac{1}{128}u_{3x}^{2}+\frac{7}{64}uu_{xx}^{2}-\frac {35%
}{64}u^{2}u_{x}^{2}+\frac{7}{64}u^{5}, \\
& \vdots
\end{split}%
\end{equation*}
As $u$ belongs to the whole hierarchy \eqref{h0} we can consider it as depending on infinitely
many evolution parameters $t_{i}$ and one spatial variable $x$:
$u=u(x,t_{1},t_{2},t_{3},...)$.

\subsection{Lax representation}

It well know that the hierarchy (\ref{h0}) can be reconstructed from an
isospectral problem. Consider the following pair of a spectral problem and its auxiliary problem
\begin{equation}
\begin{split}
L\psi & =\lambda\psi,\qquad\lambda_{t_{n}}=0, \\
\psi_{t_{n}} & =B_{n}\psi,\qquad n=1,2,\ldots,
\end{split}
\label{L1}
\end{equation}
where $L$ and $B_{n}$ are some differential operators. The compatibility
conditions for (\ref{L1}) take the form
\begin{equation}
L_{t_{n}}=[B_{n},L],\qquad n=1,2,\ldots,  \label{h1}
\end{equation}
known as the isospectral deformation equations, since the eigenvalues of the
operator $L$ are independent of all times $t_{i}$. The equations (\ref{h1})
are equivalent with the evolutionary hierarchy of PDE's \eqref{h0}. For the KdV hierarchy
\begin{equation}
L=\partial_{x}^{2}+u,\qquad B_{n}\equiv\left( L^{n-\frac{1}{2}}\right)
_{\geq0}=\sum_{i=0}^{n-1}\Bigl ( -\frac{1}{4}(\gamma_{i})_{x}+\frac{1}{2}%
\gamma_{i}\partial_{x}\Bigr ) L^{n-i-1},\qquad n=1,2,\ldots,  \label{L1a}
\end{equation}
where in particular
\begin{equation*}
\begin{split}
B_{1} & =\partial_{x}, \\
B_{2} & =\partial_{x}^{3}+\frac{3}{2}u\partial_{x}+\frac{3}{4}u_{x}, \\
B_{3} & =\partial_{x}^{5}+\frac{5}{2}u\partial_{x}^{3}+\frac{15}{4}%
u_{x}^{2}\partial_{x}^{2}+\frac{5}{8}(3u^{2}+5u_{xx})\partial_{x} +\frac{15}{%
16}(u_{3x}+2uu_{x}), \\
& \vdots
\end{split}%
\end{equation*}

As a consequence of \eqref{L1a} we can represent the linear problem
\eqref{L1} by means of polynomials in the spectral variable $\lambda$:
\begin{subequations}  \label{L2}
\begin{align}
\psi_{xx} & =\lambda\psi-u\psi, \\
\psi_{t_{n}} & =P_{n}\psi_{x}-\frac{1}{2}\left( P_{n}\right) _{x}\psi,\qquad
n=1,2,\ldots,\label{L2b}
\end{align}
\end{subequations}
where
\begin{equation}  \label{pn}
P_{n}\equiv\frac{1}{2}\sum_{i=0}^{n-1}\gamma_{i}\lambda^{n-i-1}.
\end{equation}
Then, the compatibility conditions $\left( \psi_{xx}\right) _{t_{n}}=\left(
\psi_{t_{n}}\right) _{xx}$ of the equations \eqref{L2} provide the hierarchy
\eqref{h1} in the form
\begin{equation}
u_{t_{n}}=2\left( P_{n}\right) _{x}(u-\lambda)+u_{x}P_{n}+\frac{1}{2}\left(
P_{n}\right) _{3x}\equiv\mathcal{K}_{n},\qquad n=1,2,\ldots\,.  \label{L3}
\end{equation}
The consistency of the KdV hierarchy causes that all the $\lambda$ terms in
\eqref{L3} mutually cancel.

The bi-Hamiltonian chain for the KdV hierarchy~\eqref{h0}, on the level of
co-symmetries, takes the form $\pi_{\lambda}P_{\lambda}=0$ or, explicitly
\begin{equation}
\left( P_{\lambda
}\right) _{x}(u-\lambda)+\frac{1}{2}u_{x}P_{\lambda}+\frac{1}{4}\left(
P_{\lambda}\right) _{3x}=0,  \label{L4}
\end{equation}
where
\begin{equation*}
P_{\lambda}\equiv\sum_{i=0}^{\infty}\gamma_{i}\lambda^{-i-1}
\end{equation*}
lies in the kernel of the Poisson pencil $\pi_{\lambda}\equiv\pi_{1}-\lambda%
\pi_{0}$. In fact, we can integrate \eqref{L4} to the equation
\begin{equation}
-\frac{1}{2}P_{\lambda}\left( P_{\lambda}\right) _{xx}+\frac{1}{4}\left(
P_{\lambda}\right) _{x}^{\,2}-(u-\lambda)P_{\lambda}^{\,2}=C(\lambda
)\equiv4\lambda^{-1},  \label{L4b}
\end{equation}
where $C(\lambda)$ is an arbitrary function of $\lambda$ with coefficients
being constants of integration appearing in the recursion \eqref{h2}. Here we make the simplest possible choice $C(\lambda)\equiv4\lambda^{-1}$. Solving recursively \eqref{L4b} for
coefficients of $P_{\lambda}$ one finds that $\gamma_{0}=2$, $\gamma_{1} = u$
and
\begin{equation}  \label{L4c}
\gamma_{k} = \frac{1}{16}\sum_{i=1}^{k-1} \bigl [2\gamma
_{k-i-1}(\gamma_{i})_{xx} - (\gamma_{k-i-1})_{x}(\gamma_{i})_{x}
-4\gamma_{k-i} \gamma_{i} \bigr ] + \frac{1}{4}\sum_{i=0}^{k-1} u
\gamma_{k-i-1}\gamma_{i},\qquad k\geq2.
\end{equation}
Now, the KdV flows can be obtained in the form \eqref{L3} taking $P_{n}=%
\frac{1}{2}\left[ \lambda^{n}P_{\lambda}\right] _{+}$, where $[\cdot]_{+}$
means, here, the projection on the polynomial part in $\lambda$. The
algebraic recursion formula \eqref{L4c}, for the construction of
co-symmetries $\gamma_{i}$, was originally obtained in \cite{alber79}. Let us
note that contrary to the original recursion \eqref{h2} the formula
\eqref{L4c} does not require integration.

\subsection{Zero-curvature representation}

The hierarchy (\ref{h1}) can also be reconstructed form the
so-called zero-curvature equations, which are more suitable for our further
considerations. Introducing the vector eigenfunction $\Psi=(\psi,%
\psi_{x})^{T}$ we can rewrite the linear problem for the KdV hierarchy
\eqref{L1}, or equivalently \eqref{L2}, in the form
\begin{equation}
\Psi_{t_{n}} =\mathbb{V}_{n}\Psi,\qquad n=1,2,\ldots,
\label{L5}
\end{equation}
where
\begin{equation}  \label{L6}
\mathbb{V}_{n}=
\begin{pmatrix}
-\frac{1}{2}\left( P_{n}\right)_{x} & P_{n} \\
P_{n}(\lambda-u)-\frac{1}{2}\left(P_{n}\right) _{xx} & \frac{1}{2}\left(
P_{n}\right) _{x}%
\end{pmatrix}
,\qquad n=1,2,\ldots\,.
\end{equation}
In particular
\begin{subequations}
\label{L7}
\begin{equation}
\mathbb{V}_{1}=
\begin{pmatrix}
0 & 1 \\
\lambda-u & 0%
\end{pmatrix}
,\qquad\mathbb{V}_{2}=
\begin{pmatrix}
-\frac{1}{4}u_{x} & \lambda+\frac{1}{2}u \\
\lambda^{2}-\frac{1}{2}u\lambda-\frac{1}{2}u^{2}-\frac{1}{4}u_{xx} & \frac {1%
}{4}u_{x}%
\end{pmatrix}
\label{L7a}
\end{equation}
and
\begin{equation}  \label{L7b}
\mathbb{V}_{3}=
\begin{pmatrix}
-\frac{1}{4}u_{x}\lambda-\frac{1}{16}(u_{3x}+6uu_{x}) & \lambda^{2}+\frac {1%
}{2}u\lambda+\frac{1}{8}(u_{xx}+3u^{2}) \\
\lambda^{3}-\frac{1}{2}u\lambda^{2}-\frac{1}{8}(u_{xx}+u^{2})\lambda-(\frac {%
1}{16}u_{4x}+\frac{1}{2}uu_{xx}+\frac{3}{8}u_{x}^{2}+\frac{3}{8}u^{3}) &
\frac{1}{4}u_{x}\lambda+\frac{1}{16}(u_{3x}+6uu_{x})%
\end{pmatrix}
.
\end{equation}
\end{subequations}

From the equation \eqref{L2b} it follows that $\psi_{t_1} = \psi_x$ and hence \eqref{L5}
for $n=1$ specifies to
\begin{equation}
\Psi_{x} =\mathbb{V}_{1}\Psi.
\label{L5b}
\end{equation}

The compatibility conditions $(\Psi_{x})_{t_{n}} = (\Psi_{t_{n}})_{x}$
between \eqref{L5} and \eqref{L5b} take the form of the following zero-curvature equations
\begin{equation}  \label{L9i}
\frac{d}{dt_{n}}\mathbb{V}_{1} = [\mathbb{V}_{n},\mathbb{V}_{1}] +\frac{d}{dx%
}\mathbb{V}_{n},\qquad n=1,2,\ldots\,,
\end{equation}
which are equivalent to the respective members of the KdV hierarchy~\eqref{L3}.
Here, $\frac{d}{dx}$ and $\frac{d}{dt_{n}}$ means the total
derivatives with respect to spatial $x$ and evolution $t_{n}$ variables. The
remaining zero-curvature equations coming from the conditions
$(\Psi_{t_{m}})_{t_{k}} = (\Psi_{t_{k}})_{t_{m}}$,
\begin{equation}
\frac{d}{dt_{k}}\mathbb{V}_{r}-\frac{d}{dt_{r}}\mathbb{V}_{k}+[\mathbb{V}%
_{r},\mathbb{V}_{k}]=0, \qquad r,k=1,2,\ldots,  \label{L9ii}
\end{equation}
are identically satisfied due to the commutativity of all the vector fields of the KdV hierarchy~\eqref{h0}.

\subsection{Stationary systems}

The $(n+1)$-st stationary flow is determined by the following restriction on
the $(n+1)$-st KdV symmetry:
\begin{equation}
u_{t_{n+1}}=0\qquad\text{or equivalently}\qquad\mathcal{K}_{n+1}=0,
\label{L9}
\end{equation}
which can be obtained by imposing on the linear problems \eqref{L5} the
constraint
\begin{subequations}
\label{L9ab}
\begin{equation}
\Psi_{t_{n+1}}=\lambda^{m}\mu\Psi  \label{L9a}
\end{equation}
or equivalently
\begin{equation}
\mathbb{V}_{n+1}\Psi=\lambda^{m}\mu\Psi.  \label{L9b}
\end{equation}
The factor $\lambda^{m}$ in \eqref{L9ab} is a matter of later convenience.
Indeed, the constraint \eqref{L9a} and the compatibility condition $(\Psi
_{x})_{t_{n+1}}=(\Psi_{t_{n+1}})_{x}$ gives
\end{subequations}
\begin{equation}
\frac{d}{dt_{n+1}}\mathbb{V}_{1}=0,  \label{L9cc}
\end{equation}
which is equivalent to \eqref{L9}, or alternatively the compatibility
condition between eigenvalue problem \eqref{L9b} and $\Psi_{x}=\mathbb{V}%
_{1}\Psi$ yields the Lax equation
\begin{equation*}
\frac{d}{dx}\mathbb{V}_{n+1}=[\mathbb{V}_{1},\mathbb{V}_{n+1}],
\end{equation*}
which combined with the zero-curvature equation \eqref{L9i} for $k=n+1$
gives again \eqref{L9cc}.

The differential order of $(n+1)$-st vector field $\mathcal{K}_{n+1}$ is
equal to $2n+1$, which means that the vector field $\mathcal{K}_{n+1}$ depends
on $2n+2$ jet variables: $u,u_{x},\ldots,u_{(2n+1)x}$. The stationary
restriction \eqref{L9} provides constraint on the infinite-dimensional
(functional) manifold, on which the KdV hierarchy is defined, reducing it to
the finite-dimensional (stationary) submanifold $\mathcal{M}_{n}$ of dimension $(2n+1)$.
Using \eqref{L9} and its differential consequences we can
eliminate all terms of order $2n+1$ and higher. Thus, the coordinates on the
stationary manifold $\mathcal{M}_{n}$ are provided by the jet coordinates:
$u,u_{x},\ldots,u_{(2n)x}$. Due to the integrability the constraint is
invariant with respect to all flows from the KdV hierarchy. As result the
infinite hierarchy (\ref{h0}) reduces to the finite system:
\begin{equation}
u_{t_{1}}=\mathcal{K}_{1},\qquad u_{t_{2}}=\mathcal{K}_{2},\qquad\ldots,%
\qquad u_{t_{n}}=\mathcal{K}_{n},\qquad\mathcal{K}_{n+1}=0  \label{L10a}
\end{equation}
which further will be called the \emph{$n$-th stationary KdV system}.

The finite hierarchy of associated Lax equations is given by equations
\begin{equation}
\frac{d}{dt_{k}}\mathbb{V}_{n+1}=[\mathbb{V}_{k},\mathbb{V}_{n+1}],\qquad
k=1,2,\ldots,n,  \label{L10b}
\end{equation}
valid under the constraint \eqref{L9}. Notice that from
\eqref{L9cc} or directly \eqref{L9} it follows that
$(\mathbb{V}_{k})_{t_{n+1}}=0$. Thus, one obtains the Lax
equations \eqref{L10b} simply imposing \eqref{L9cc} on the
zero-curvature equations \eqref{L9ii}, with $r=n+1$, or by the
compatibility conditions between eigenvalue problem \eqref{L9b}
and the respective linear problems \eqref{L5}.

After imposing the constraint \eqref{L9b} the existence of nontrivial
solutions for the respective linear problems enforces the characteristic
equation
\begin{equation}
\det\left( \mathbb{V}_{n+1}-\lambda^{m}\mu\mathbb{I}\right) =0,  \label{L11}
\end{equation}
associated with \eqref{L9}. Equation \eqref{L11} determines the spectral
curve
\begin{equation}
-\frac{1}{2}P_{n+1}\left( P_{n+1}\right) _{xx}+\frac{1}{4}\left(
P_{n+1}\right) _{x}^{\,2}-(u-\lambda)P_{n+1}^{\,2}=\lambda^{2m}\mu ^{2},
\label{L12}
\end{equation}
which takes the more explicit form
\begin{equation}
\lambda^{2n+1}+\sum_{k=0}^{n}h_{k}\lambda^{n-k}=\lambda^{2m}\mu ^{2},
\label{L13}
\end{equation}
where
\begin{equation*}
h_{k} = -\frac{1}{16}\sum_{i=0}^{n-k} \bigl [2\gamma_{n-i}(\gamma
_{i+k})_{xx} -(\gamma_{n-i})_{x}(\gamma_{i+k})_{x} + 4 u
\gamma_{n-i}\gamma_{i+k} \bigr ] + \frac{1}{4} \sum_{i=1}^{n-k}%
\gamma_{n-i+1}\gamma_{i+k}.
\end{equation*}
In particular, by \eqref{L4c}
\begin{subequations}
\label{L14}
\begin{equation}  \label{L14a}
h_{0} = -\frac{1}{16}\sum_{i=0}^{n} \bigl [2\gamma_{n-i}(\gamma_{i})_{xx}
-(\gamma_{n-i})_{x}(\gamma_{i})_{x} + 4 u \gamma_{n-i}\gamma_{i} \bigr ] +
\frac{1}{4} \sum_{i=1}^{n}\gamma_{n-i+1}\gamma_{i} \equiv-\gamma_{n+1}
\end{equation}
and
\begin{equation}  \label{L14b}
h_{n}=-\frac{1}{8}\gamma_{n}(\gamma_{n})_{xx}+\frac{1}{16}%
(\gamma_{n})_{x}^{2}-\frac{1}{4}u\gamma_{n}^{2}.
\end{equation}
\end{subequations}
In fact, the coefficients $h_{0},\ldots,h_{n}$ of \eqref{L13}
are constants of motion of the respective stationary system
\eqref{L10a}, and thus the spectral curve \eqref{L13} describes a
common level of them.

\begin{remark}
Observe that the l.h.s.~of the spectral curve \eqref{L12} could be obtained
alternatively as follows. By \eqref{L3} the $(n+1)$-st stationary flow
\eqref{L9} is given by the condition
\begin{equation}
2\left( P_{n+1}\right) _{x}(u-\lambda)+u_{x}P_{n+1}+\frac{1}{2}\left(
P_{n+1}\right) _{3x}=0,  \label{ff1}
\end{equation}
which can be directly integrated to the form:
\begin{equation}
-\frac{1}{2}P_{n+1}\left( P_{n+1}\right) _{xx}+\frac{1}{4}\left(
P_{n+1}\right) _{x}^{\,2}-(u-\lambda)P_{n+1}^{\,2}=C(\lambda).  \label{ff2}
\end{equation}
Here $C(\lambda)=\lambda^{2n+1}+\sum_{k=0}^{n}\varepsilon_{k}\lambda^{n-k}$
is an 'integral' sum in $\lambda$ with constant coefficients $\varepsilon
_{i}=h_{i}$. Thus, differentiating the spectral curve \eqref{ff2} or
\eqref{L12} one reconstructs the stationary condition \eqref{ff1}. The
equation \eqref{ff2} plays a crucial role in the construction presented in
the article \cite{alber81}.
\end{remark}

\subsection{Two integrated representations of stationary KdV systems}

Since the KdV hierarchy is bi-Hamiltonian, the $(n+1)$-st stationary flow \eqref{L9} can be written in two ways:
\begin{equation*}
\mathcal{K}_{n+1}=\pi_{0}\gamma_{n+1}=\pi_{1}\gamma_{n}=0,
\end{equation*}
which leads to two different integrated representations of the $n$-th stationary KdV system \eqref{L10a}.
Indeed, integrating the first Hamiltonian structure, $\pi_{0}\gamma_{n+1}=0$ we find that
\begin{equation}
\gamma_{n+1}+c=0,  \label{L17}
\end{equation}
where $c$ is an integration constant. This means that the constraint
\eqref{L17} defines a (Hamiltonian) foliation of the stationary manifold
$\mathcal{M}_{n}$, of codimension $1$, parameterized by the constant $c$:
\begin{equation*}
\mathcal{M}_{n}=\bigcup_{c\in\mathbb{R}}\mathcal{M}_{n,c}.
\end{equation*}
In result, the first integrated representation of the $n$-th stationary KdV system is
given by
\begin{equation}
u_{t_{1}}=\mathcal{K}_{1},\qquad u_{t_{2}}=\mathcal{K}_{2},\qquad\ldots,%
\qquad u_{t_{n}}=\mathcal{K}_{n},\qquad\gamma_{n+1}+c=0,  \label{L17a}
\end{equation}
which constitutes a system of $n$ ODE's on the $2n$-dimensional leaf
$\mathcal{M}_{n,c}$ endowed with the jet
coordinates $u,u_{x},\ldots,u_{(2n-1)x}$ (the higher derivatives of
$u$ with respect to $x$ are not needed as they can all be eliminated by the differential constraint
\eqref{L17}). The associated Lax equations are then given by
\eqref{L10b} with the imposed constraint \eqref{L17}. In this case the spectral
curve \eqref{L13}, choosing $m=0$ and taking into account \eqref{L14a},
takes the form
\begin{equation}
\lambda^{2n+1}+c\lambda^{n}+\sum_{k=1}^{n}H_{k}\lambda^{n-k}=\mu ^{2},
\label{L17b}
\end{equation}
where $H_k:=h_k$.

Integrating the second Hamiltonian structure, $\pi_{1}\gamma_{n}=0$, we find
another constraint
\begin{equation}
\frac{1}{2}\gamma_{n}(\gamma_{n})_{xx}-\frac{1}{4}(\gamma_{n})_{x}^{2}+u%
\gamma_{n}^{2}+ 4\bar{c}=0  \label{L18}
\end{equation}
which defines an alternative foliation of the stationary manifold $\mathcal{%
M}_{n}$ parameterized by the constant $\bar{c}$. Thus, \eqref{L18}
provides the second representation of the $n$-th stationary KdV system
\eqref{L10a}:
\begin{equation}
u_{t_{1}}=\mathcal{K}_{1},\qquad u_{t_{2}}=\mathcal{K}_{2},\qquad\ldots,%
\qquad u_{t_{n}}=\mathcal{K}_{n},\qquad\frac{1}{2}\gamma_{n}(%
\gamma_{n})_{xx}-\frac{1}{4}(\gamma_{n})_{x}^{2}+u\gamma_{n}^{2}+4\bar{c}=0
\label{L18a}
\end{equation}
on the $2n$-dimensional leaf $\bar{\mathcal{M}}_{n,\bar{c}}$
endowed with the same set of jet coordinates
$u,u_{x},\ldots,u_{(2n-1)x}$ (again, higher order derivatives can
be eliminated by \eqref{L18}). The Lax representation of the
system \eqref{L18a} is again given by \eqref{L10b} but now with
the imposed condition \eqref{L18}. The spectral curve \eqref{L13},
taking into account \eqref{L14b} and choosing $m=1$, attains now
the form
\begin{equation}
\lambda^{2n}+\bar{c}\lambda^{-1}+\sum_{k=1}^{n}\bar{H}_{k}\lambda^{n-k}=\lambda
\mu^{2},  \label{L18b}
\end{equation}
where $\bar{H}_k:= h_{k-1}$.

Note that the two foliations of the stationary manifold $\mathcal{M}_{n}$,
defined by the constraints \eqref{L17} and \eqref{L18}, are not equivalent,
and one can show that they are mutually transversal.

\begin{remark}
The two integrated representations of (the same) stationary KdV system, as given by \eqref{L17a} and by
\eqref{L18a}, are not equivalent because the complete form of the
evolution equations is only given in jet coordinates on the leaves
$\mathcal{M}_{n,c}$ and $\bar{\mathcal{M}}_{n,\bar{c}}$ by
taking into account the respective constraints \eqref{L17} and
\eqref{L18} and their differential consequences. However, both
representations become equivalent when considered on the extended
$(2n+1)$-dimensional phase space, the stationary manifold
$\mathcal{M}_n$. The equivalence is given by the appropriate Miura
map, see Lemma~\ref{thm2}.
\end{remark}

\begin{example}
The first representation \eqref{L17a} of the stationary KdV system, given
for $n=2$, is constituted by the first two flows from the KdV hierarchy:
\begin{subequations}
\label{E1}
\begin{equation}
u_{t_{1}}=u_{x}\equiv K_{1},\qquad u_{t_{2}}=\frac{1}{4}u_{xxx}+\frac{3}{2}%
uu_{x}\equiv\mathcal{K}_{2},  \label{E1a}
\end{equation}
and the constraint
\begin{equation}
\frac{1}{16}u_{4x}+\frac{5}{8}u u_{xx}+\frac{5}{16}u_{x}^{2}+\frac{5}{8}%
u^{3}+c=\gamma_{3}+c=0.  \label{E1b}
\end{equation}
The Lax representation of the stationary system \eqref{E1} is given by the
Lax equations
\end{subequations}
\begin{equation*}
\frac{d}{dt_{1}}\mathbb{V}_{3}=[\mathbb{V}_{1},\mathbb{V}_{3}],\qquad\frac {d%
}{dt_{2}}\mathbb{V}_{3}=[\mathbb{V}_{2},\mathbb{V}_{3}],
\end{equation*}
where the matrices $\mathbb{V}_{1}$ and $\mathbb{V}_{2}$ are given by
\eqref{L7a} and the matrix $\mathbb{V}_{3}$ \eqref{L7b} under the constraint
\eqref{E1b} takes the form
\begin{equation}
V_{3}=%
\begin{pmatrix}
-\frac{1}{4}u_{x}\lambda-\frac{1}{16}(u_{3x}+6u u_{x}) & \lambda^{2}+\frac {1%
}{2}u\lambda+\frac{1}{8}(u_{xx}+3u^{2}) \\
\lambda^{3}-\frac{1}{2}u\lambda^{2}-\frac{1}{8}(u_{xx}+u^{2})\lambda+\frac {1%
}{8}u u_{xx}-\frac{1}{16}u_{x}^{2}+\frac{1}{4}u^{3}+c & \frac{1}{4}%
u_{x}\lambda+\frac{1}{16}(u_{3x}+6u u_{x})%
\end{pmatrix}
.  \label{E2}
\end{equation}
The associated spectral curve \eqref{L17b} for $n=2$ is given by
\begin{equation*}
\lambda^{5}+c\lambda^{2}+H_{1}\lambda+H_{2}=\mu^{2},
\end{equation*}
where one finds the following (nontrivial) integrals of motion:
\begin{equation}  \label{E2h}
\begin{split}
H_{1} & = \frac{1}{32}u_{x}u_{3x}-\frac{1}{64}u_{xx}^{2}+\frac{5}{32}u
u_{x}^{2}+\frac{5}{64}u^{4} +\frac{1}{2}cu, \\
H_{2} & =\frac{3}{64}u u_{x}u_{3x} +\frac{1}{256}u_{3x}^{2}-\frac{1}{128}%
u_{x}^{2}u_{xx} +\frac{5}{64}u^{3}u_{xx} +\frac{15}{128}u^{2}u_{x}^{2}+\frac{%
1}{64}uu_{xx}^{2}+\frac{3}{32}u^{5} + \frac{1}{8}cu_{xx} + \frac {3}{8}%
cu^{2} .
\end{split}%
\end{equation}
\end{example}

\begin{example}
The second representation \eqref{L18a} of the stationary KdV system, for $%
n=2 $, is given by the flows:
\begin{subequations}
\label{E3}
\begin{equation}
u_{t_{1}}=u_{x}\equiv K_{1},\qquad u_{t_{2}}=\frac{1}{4}u_{xxx}+\frac{3}{2}u
u_{x}\equiv\mathcal{K}_{2},  \label{E3a}
\end{equation}
and the constraint
\begin{equation}
\frac{1}{32}u_{xx}u_{4x}+\frac{3}{32}u^{2}u_{4x}-\frac{1}{64}u_{3x}^{2}-
\frac{3}{16}u u_{x}u_{3x}+\frac{3}{16}u_{x}^{2}u_{xx}+\frac{1}{4}
u
u_{xx}^{2}+\frac{15}{16}u^{3}u_{xx}+\frac{9}{16}u^{5}+4\bar{c}=0.
\label{E3b}
\end{equation}

The Lax representation of the stationary system \eqref{E3} is given by
\end{subequations}
\begin{equation*}
\frac{d}{dt_{1}}\mathbb{V}_{3}=[\mathbb{V}_{1},\mathbb{V}_{3}],\qquad\frac {d%
}{dt_{2}}\mathbb{V}_{3}=[\mathbb{V}_{2},\mathbb{V}_{3}],
\end{equation*}
where the matrices $\mathbb{V}_{1}$ and $\mathbb{V}_{2}$ are given by %
\eqref{L7a} and the matrix $\mathbb{V}_{3}$ \eqref{L7b} under the constraint %
\eqref{E3b} takes the form
\begin{equation}
\mathbb{V}_{3}=%
\begin{pmatrix}
-\frac{1}{4}u_{x}\lambda-\frac{1}{16}(u_{3x}+6uu_{x}) & \lambda^{2}+\frac {1%
}{2}u\lambda+\frac{1}{8}(u_{xx}+3u^{2}) \\
\lambda^{3}-\frac{1}{2}u\lambda^{2}-\frac{1}{8}(u_{xx}+u^{2})\lambda + *
& \frac{1}{4}u_{x}\lambda+\frac{1}{16}(u_{3x}+6uu_{x})%
\end{pmatrix}
,  \label{E4}
\end{equation}
where
\begin{equation}
*=-\frac{u_{3x}^{2}+12uu_{x}u_{3x}+36u^{2}u_{x}^{2}-256\bar{c}}{
32u_{2x}+ 96u^{2} }.
\end{equation}

The associated spectral curve \eqref{L18b} for $n=2$ is given by
\begin{equation*}
\lambda^{4}+\bar{c}\lambda^{-1}+\bar{H}_{1}\lambda+\bar{H}_{2}=\lambda\mu^{2}
\end{equation*}
with the integrals of motion:
\begin{equation}  \label{E4h}
\begin{split}
\bar{H}_{1}& =-\frac{u_{3x}^{2}-2u_{x}^{2}u_{2x}+12uu_{x}u_{3x}+30u^{2}u_{x}^{2}}{%
32(u_{2x}- 3u^{2}) } -\frac{1}{8}uu_{2x}-\frac{1}{4}u^3 +\frac{8 \bar{c}}{%
u_{2x}+3u^2} , \\
\bar{H}_{2}& =-\frac{uu_{3x}^{2}-2u_{x}u_{2x}u_{3x}
+6u^{2}u_{x}u_{3x}-12uu_{x}^{2}u_{2x}}{64(u_{2x} + 3u^{2}) } -\frac{1}{64}%
u_{2x}^2-\frac{1}{16} u^2 u_{2 x}-\frac{3}{64}u^4 +\frac{4 \bar{c}u}{%
u_{2x}+3u^2}.
\end{split}%
\end{equation}
\end{example}

\section{St\"{a}ckel systems
\label{3}}

In this chapter we gather the necessary information about St\"{a}ckel systems.

\subsection{St\"{a}ckel systems in separation coordinates}

Let us consider the spectral curve \cite{skl} in the form
\begin{equation}
\sigma(\lambda)+\sum_{k=1}^{n}H_{k}\lambda^{n-k}=\lambda^{m}\mu^{2},\qquad
m\in\mathbb{Z},  \label{S1}
\end{equation}
where $\sigma(\lambda)$ is a (Laurent) polynomial in the variables
$\lambda$ and $\lambda^{-1}$. The associated separable systems
belong to the so-called Benenti subclass of St\"{a}ckel systems
\cite{ben1,ben2,blaszak2006,blaszak2007}. The separation relations
are reconstructed by taking $n$ copies of \eqref{S1} with respect
to the coordinates $(\bm{\lambda},\bm{\mu})$ on a phase space
$M=T^{\ast}Q$, where
$\bm{\lambda}=(\lambda_{1},\ldots,\lambda_{n})^{T}$ are local
coordinates on the configuration space $Q$ and
$\bm{\mu}=(\mu_{1},\ldots,\mu_{n})^{T}$ are the (fibre) momentum
coordinates. Thus, solving the linear system
\begin{equation*}
\sigma(\lambda_{i})+\sum_{k=1}^{n}H_{k}\lambda_{i}^{n-k}=\lambda_{i}^{m}%
\mu_{i}^{2},\qquad i=1,\ldots,n,
\end{equation*}
with respect to functions $H_{k}=H_{k}(\bm{\lambda},\bm{\mu})$ we obtain $n$
quadratic in momenta Hamiltonians on $M$
\begin{equation}
H_{k}=\frac{1}{2}\bm{\mu}^{T}K_{k}G_{m}\bm{\mu}+V_{k},\qquad k=1,\ldots ,n,
\label{S3}
\end{equation}
where $G_{m}$ represents the contravariant metric, defined by the first
Hamiltonian $H_{1}$, on the configuration space $Q$. In fact
\begin{equation*}
G_{m}=L^{m}G_{0},\qquad G_{0}=2\,\mathrm{diag}\left( \frac{1}{\Delta_{1}}%
,\ldots,\frac{1}{\Delta_{n}}\right) ,\qquad\Delta_{i}=\prod_{j\neq
i}(\lambda_{i}-\lambda_{j}).
\end{equation*}
Here, $K_{k}$ are respective Killing tensors and $L$ is a special conformal
Killing tensor \cite{C2001}, given by:
\begin{equation*}
K_{k}=(-1)^{k+1}\mathrm{diag}\left( \frac{\partial s_{k}}{\partial\lambda
_{1}},\ldots,\frac{\partial s_{k}}{\partial\lambda_{n}}\right) ,\qquad L=%
\mathrm{diag}(\lambda_{1},\ldots,\lambda_{n}),
\end{equation*}
where $s_{k}$ are the elementary symmetric polynomials in $\lambda_{i}$. The
potential functions $V_{k}$ are given by
\begin{equation}  \label{S6}
V_{k}=(-1)^{k+1}\sum_{i=1}^{n}\frac{\partial s_{k}}{\partial \lambda_{i}}%
\frac{\sigma(\lambda_{i})}{\Delta_{i}}.
\end{equation}

The Hamiltonians \eqref{S3} are in involution with respect to the
Poisson bracket defined by
\begin{equation*}
\{\cdot,\cdot\}=\sum_{i=1}^{n}\frac{\partial}{\partial\lambda_{i}}\wedge
\frac{\partial}{\partial\mu_{i}}
\end{equation*}
and moreover, by the very construction, they are all separable in the variables  $(\bm{\lambda},\bm{\mu})$.
The evolution of any observable $\xi$ with respect to the Hamiltonian $H_{k}$
has the form $\xi_{t_{k}}=\{\xi,H_{k}\} $ and the Hamiltonian evolution
equations are
\begin{equation}
\bm{\lambda}_{t_{k}}=\{\bm{\lambda},H_{k}\},\qquad\bm{\mu}_{t_{k}}=\{\bm{\mu}%
,H_{k}\},\qquad k=1,\ldots,n.  \label{S9}
\end{equation}

\subsection{Lax representation}

As it shown in \cite{blaszak2019}, the Hamiltonian evolution equations
\eqref{S9} associated with the spectral curves \eqref{S1} can be represented
by the (isospectral) Lax equations
\begin{equation}
\frac{d}{dt_{k}}\mathbb{L}=\left[ \mathbb{U}_{k},\mathbb{L}\right] ,\qquad
k=1,\ldots,n,  \label{T1}
\end{equation}
with $\mathbb{L}$ and $\mathbb{U}_{k}$ being $2\times2$ traceless
matrices depending rationally on the spectral parameter~$\lambda$.
The Lax matrix $\mathbb{L}$ has the form
\begin{equation}
\mathbb{L}=%
\begin{pmatrix}
\bm{v} & \bm{u} \\
\bm{w} & -\bm{v}%
\end{pmatrix}
,  \label{T2}
\end{equation}
where in the separation coordinates $(\bm{\lambda},\bm{\mu})$ the
entries are\footnote{In the construction of the Lax equations in
\cite{blaszak2019} there is some freedom. In the present article
we choose, using for a moment the notation from
\cite{blaszak2019},
$g(\lambda)=\frac{1}{2}f(\lambda)=\lambda^{m}$.}
\begin{equation*}
\bm{u}=\prod_{k=1}^{n}(\lambda-\lambda_{k})\equiv\lambda^{n}+%
\sum_{k=1}^{n}(-1)^{k}s_{k}\lambda^{n-k},
\end{equation*}%
\begin{equation*}
\bm{v}=\sum_{k=1}^{n}(-1)^{k+1}\left[ \sum_{i=1}^{n}\frac{\partial s_{k}}{%
\partial\lambda_{i}}\frac{\lambda_{i}^{m}\mu_{i}}{\Delta_{i}}\right]
\lambda^{n-k}
\end{equation*}
and
\begin{equation}
\bm{w}=\frac{1}{\bm{u}}\biggl [\lambda^{m}\Bigl(\sigma(\lambda)+\sum
_{k=1}^{n}H_{k}\lambda^{n-k}\Bigr)-\bm{v}^{2}\biggr ].  \label{T3c}
\end{equation}
In fact, $\bm{w}$ is defined so that the spectral curve \eqref{S1} can be
reconstructed from the characteristic equation for $\mathbb{L}$, since
\begin{equation*}
0=\det\bigl[\mathbb{L}-\lambda^{m}\mu\mathbb{I}\bigr]=
-\lambda^{m}\Bigl(\sigma(\lambda)+\sum
_{k=1}^{n}H_{k}\lambda^{n-k} -\lambda^{m}\mu^{2}\Bigr).
\end{equation*}
One can show that the expression in the quadratic bracket in \eqref{T3c}
factorizes so that $\bm{w}$ takes the form of a Laurent polynomial in
$\lambda$:
\begin{equation}
\bm{w}=\lambda^{m}\left[ \frac{\sigma(\lambda)-\lambda^{-m}\bm{v}^{2}}{\bm{u}%
}\right] _{+}.  \label{T5}
\end{equation}
Here, the operation $[\cdot]_{+}$ means the projection on the uniquely
defined quotient of the division of an analytic function $A$ over a (pure)
polynomial $\bm{u}$ such that the following decomposition holds:
\begin{equation*}
A = \left[ \frac{A}{\bm{u}}\right] _{+} \bm{u} + r,
\end{equation*}
where the (unique) remainder $r$ is a lower degree polynomial than the
polynomial $\bm{u}$, see for details \cite{blaszak2019}. In particular when $%
A$ is a Laurent polynomial we have
\begin{equation*}
\left[ \frac{A}{\bm{u}}\right] _{+}\equiv\left[ \frac{[A]_{\geqslant0}}{%
\bm{u}}\right] _{\geqslant0}+\left[ \frac{[A]_{<0}}{\bm{u}}\right] _{<0},
\end{equation*}
where $[\cdot]_{\geqslant0}$ is the projection on the part consisting of
non-negative degree terms in the expansion into Laurent series at $\infty$
and $[\cdot]_{<0}$ is the projection on the part consisting of negative
degree terms in the expansion into Laurent series at $0$.

Further, the generating matrices $\mathbb{U}_{k}$ are defined by
\begin{equation}  \label{T7}
\mathbb{U}_{k} := \left[ \frac{\bm{u}_{k}\mathbb{L}}{\bm{u}} \right] _{+}
\equiv%
\begin{pmatrix}
\left[ \frac{\bm{u}_{k}\bm{v}}{\bm{u}} \right] _{+} & \bm{u}_{k} \\
\left[ \frac{\bm{u}_{k} \bm{w}}{\bm{u}}\right] _{+} & -\left[ \frac {\bm{u}%
_{k}\bm{v}}{\bm{u}} \right] _{+}%
\end{pmatrix}
,\qquad k=1,\ldots,n,
\end{equation}
where
\begin{equation*}
\bm{u}_{k} := \left[ \frac{\bm{u}}{\lambda^{n-k+1}}\right] _{+} \equiv
\lambda^{k-1}+\sum_{i=1}^{k-1}(-1)^{k}s_{k}\lambda^{k-i-1}.
\end{equation*}

The evolution of the Lax matrix \eqref{T2} with respect to Hamiltonian
equations \eqref{S9}, and consequently the Lax equations \eqref{T1}, can be
directly derived from the following useful relations:
\begin{subequations}
\label{TH}
\begin{align*}
\left\{ \bm{u},H_{k}\right\} & =-2\bm{u}_{k}\bm{v}+2\bm{u}\left[ \frac{\bm{u}%
_{k}\bm{v}}{\bm{u}}\right] _{+}, \\
\left\{ \bm{v},H_{k}\right\} & =\bm{u}_{k}\bm{w}-\bm{u}\left[ \frac{\bm{u}%
_{k}\bm{w}}{\bm{u}}\right] _{+}, \\
\left\{ \bm{w},H_{k}\right\} & =-2\bm{w}\left[ \frac{\bm{u}_{k}\bm{v}}{\bm{u}%
}\right] _{+}+2\bm{v}\left[ \frac{\bm{u}_{k}\bm{w}}{\bm{u}}\right] _{+},
\end{align*}%
which were obtained in \cite{blaszak2019}. Moreover, considering \eqref{TH} for $k=1$ and observing that $\bm{u}_{1}=1$
and $\left[ \frac{\bm{v}}{\bm{u}}\right] _{+} = 0$ we can rewrite the
matrices \eqref{T2} and \eqref{T7} in the form:
\end{subequations}
\begin{subequations}
\label{T112}
\begin{equation}  \label{T11}
\mathbb{L} =
\begin{pmatrix}
-\frac{1}{2}\dot{\bm{u}} & \bm{u} \\
-\frac{1}{2}\ddot{\bm{u}} + \bm{u} Q & \frac{1}{2}\dot{\bm{u}}%
\end{pmatrix}%
\end{equation}
and
\begin{equation}  \label{T12}
\mathbb{U}_{k} =
\begin{pmatrix}
-\frac{1}{2}\dot{\bm{u}}_{k} & \bm{u}_{k} \\
-\frac{1}{2}\ddot{\bm{u}}_{k} + \bm{u}_{k} Q & \frac{1}{2}\dot{\bm{u}}_{k}%
\end{pmatrix}%
,
\end{equation}
where $Q \equiv\left[ \frac{\bm{w}}{\bm{u}}\right] _{+}$. Here, the dot
means the derivative with respect to the first Hamiltonian flow, i.e. $\dot{%
\xi} \equiv\xi_{t_{1}}$, and one can see that $\dot{\bm{u}}_{k} = \bigl [%
\frac {\bm{u}_{k} \dot{\bm{u}}}{\bm{u}}\bigr]_{+}$. Now, the connection
between \eqref{T112} and \eqref{L6} is apparent.

\subsection{St\"{a}ckel systems in Vi\`{e}te's coordinates}

From the point of view of expressing the Hamiltonian systems \eqref{S9} in
the Lax form the most practical are the so-called Vi\`{e}te's
coordinates, defined as
\end{subequations}
\begin{equation}
q_{i}=(-1)^{i}s_{i},\qquad p_{i}=-\sum_{k=1}^{n}\frac{\lambda_{k}^{n-i}\mu
_{k}}{\Delta_{k}},\qquad i=1,\ldots,n.  \label{V1}
\end{equation}
Since the above transformation is a point transformation on $M$, these coordinates are also canonical
with respect to the same Poisson bracket $\{\cdot,\cdot\}$ which now reads
$\{\cdot,\cdot\}=\sum_{i}\frac{\partial}{\partial q_{i}}\wedge \frac{\partial}{\partial p_{i}}$.
Let $\bm{p}=(p_{1},\ldots,p_{n})^{T}$ and $\bm{q}=(q_{1},\ldots,q_{n})^{T}$.
In this coordinates the geodesic part of the Hamiltonians~$H_{k}$ is always
polynomial function of their arguments and the potentials $V_{k}$ are either
polynomials or rational functions. In Vi\`{e}te's coordinates the
Hamiltonians \eqref{S3} take the form
\begin{equation*}
H_{k}=\frac{1}{2}\bm{p}^{T}K_{k}G_{m}\bm{p}+V_{k},\qquad k=1,\ldots,n,
\end{equation*}
and the respective Hamiltonian evolution equations are
\begin{equation}
\bm{q}_{t_{k}}=\{\bm{q},H_{k}\},\qquad\bm{p}_{t_{k}}=\{\bm{p},H_{k}\},
\label{V3}.
\end{equation}

For $\sigma(\lambda)=\sum_{i}\alpha_{i}\lambda^{i}$ the potential functions
\eqref{S6} are given by $V_{k}=\sum_{i}\alpha_{i}\mathcal{V}_{k}^{(i)},$
where the so-called elementary separable potentials $\mathcal{V}_{k}^{(i)}$
can be explicitly constructed by the recursion formula \cite{blaszak2011}
\begin{equation*}
\mathcal{V}^{(i)}=R^{i}\mathcal{V}^{(0)},\qquad\mathcal{V}^{(i)}=\bigl(%
\mathcal{V}_{1}^{(i)},\ldots,\mathcal{V}_{n}^{(i)}\bigr)^{T},\qquad\mathcal{V%
}^{(0)}=(0,\ldots,0,-1)^{T},
\end{equation*}
where
\begin{equation}
R=
\begin{pmatrix}
-q_{1} & 1 & 0 & 0 \\
\vdots & 0 & \ddots & 0 \\
\vdots & 0 & 0 & 1 \\
-q_{n} & 0 & 0 & 0%
\end{pmatrix}
,\qquad R^{-1}=
\begin{pmatrix}
0 & 0 & 0 & -\frac{1}{q_{n}} \\
1 & 0 & 0 & \vdots \\
0 & \ddots & 0 & \vdots \\
0 & 0 & 1 & -\frac{q_{n-1}}{q_{n}}%
\end{pmatrix}
.  \label{V5}
\end{equation}

In Vi\`{e}te's coordinates the metric $G_{0}$ for $m=0$ has the form
\begin{equation*}
(G_{0})^{ij}=%
\begin{cases}
2q_{i+j-n-1} & \text{if}\quad i+j\geq n+1, \\
0 & \text{otherwise},
\end{cases}
\end{equation*}
 and the metrics for arbitrary $m$ are given by $G_{m}=L^{m}G_{0}$,
where in Vi\`{e}te's coordinates the special conformal Killing
tensor $L$ has the matrix representation identical to $R$
\eqref{V5} \cite{blaszak2007}. Moreover, the Killing tensors
$K_{r}$, for $r=1,\ldots,n$, are given by
\begin{equation*}
(K_{r})_{j}^{i}=%
\begin{cases}
q_{i-j+r-1} & \text{if}\quad i\leqslant j\quad\text{and}\quad
r\leqslant j,
\\
-q_{i-j+r-1} & \text{if}\quad i>j\quad\text{and}\quad r>j, \\
0 & \text{otherwise}.%
\end{cases}%
\end{equation*}
Notice that $(K_{1})_{j}^{i}=\delta_{j}^{i}$.
For convenience, in the above definitions we set $q_{0}=1$ and $q_{l}=0$ for $l<0$ or $l>n$.

In Vi\`{e}te's coordinates $\bm{u}$ in the Lax matrix
\eqref{T2} is simply given by
\begin{equation}  \label{uq}
\bm{u} = \lambda^{n} + \sum_{k=1}^{n} q_{k} \lambda^{n-k},
\end{equation}
and by simple calculation, involving the change of coordinates for the
metric $G_{m}$, and observation that $\displaystyle{\mu_{i} = \sum_{k=1}^{n}
(-1)^{k}\frac{\partial s_{k}}{\partial\lambda_{i}}p_{k}}$, one finds (again)
that $\bm{v}$ has the form
\begin{equation*}
\bm{v} = -\frac{1}{2}\sum_{k=1}^{n}\Bigl [\sum_{l=1}^{n} (G_{m})^{kl}p_{l} %
\Bigr] \lambda^{n-k} \equiv-\frac{1}{2} \dot{\bm{u}}.
\end{equation*}
Finally $\bm{w}$ can be obtained from the formula \eqref{T3c} or \eqref{T5}.

\section{St\"{a}ckel representations of KdV stationary systems \label{4}}

The first St\"{a}ckel representation of the $n$-th KdV stationary system is
given by the spectral curve \eqref{L17b},
\begin{equation}
\lambda^{2n+1}+c\lambda^{n}+\sum_{k=1}^{n}H_{k}\lambda^{n-k}=\mu ^{2},
\label{C1}
\end{equation}
which is a special case of the general case \eqref{S1} with $m=0$ and $%
\sigma(\lambda)=\lambda^{2n+1}+c\lambda^{n}$. In this case, the Hamiltonians
$H_{k}$  in the Vi\`{e}te's coordinates $(\bm{q},\bm{p})$ are given by
\begin{equation}
H_{k}=\frac{1}{2}\bm{p}^{T}K_{k}G_{0}\bm{p}+\mathcal{V}_{k}^{(2n+1)}+c%
\mathcal{V}_{k}^{(n)},\qquad k=1,\ldots,n.  \label{C2}
\end{equation}
The second representation is associated with the spectral curve \eqref{L18b}
\begin{equation}
\lambda^{2n}+\bar{c}\lambda^{-1}+\sum_{k=1}^{n}\bar{H}_{k}\lambda^{n-k}=\lambda
\mu^{2},  \label{C3}
\end{equation}
which is a special case of \eqref{S1} with $m=1$ and
$\sigma(\lambda )=\lambda^{2n}+\bar{c}\lambda^{-1}$. In this case
the Hamiltonians $\bar{H}_{k} $ are
\begin{equation}
\bar{H}_{k}=\frac{1}{2}\bm{p}^{T}K_{k}G_{1}\bm{p}+\mathcal{V}_{k}^{(2n)}+\bar{c}
\mathcal{V}_{k}^{(-1)},\qquad k=1,\ldots,n.  \label{C4}
\end{equation}
The Lax representation \eqref{T1} of respective Hamiltonian flows \eqref{V3}
are generated by the Lax operator~\eqref{T2} or equivalently \eqref{T11}
with the same, in both representations, $Q$ term:
\begin{equation}  \label{C5}
Q \equiv\left[ \frac{\bm{w}}{\bm{u}}\right] _{+} = \left[ \frac{%
\lambda^{2n+1}}{\bm{u}^{2}}\right] _{\geqslant0} = \lambda- 2q_{1}.
\end{equation}
One obtains \eqref{C5} using the formula \eqref{T5} for $\bm{w}$ and
observing that, in this particular cases, only the term with the highest
degree in $\sigma(\lambda)$ contributes to the form of $Q$.

The equivalence between the appropriate St\"{a}ckel representations and the
representations \eqref{L17a} and \eqref{L18a} of $n$-th stationary KdV
system \eqref{L10a} is apparent on the level of Lax equations \eqref{L10b}
(valid under the respective constraints) and \eqref{T1} if we make the following
identifications:
\begin{equation*}
\mathbb{L}\equiv\mathbb{V}_{n+1},\qquad\mathbb{U}_{k}\equiv\mathbb{V}%
_{k},\qquad k=1,\ldots,n,
\end{equation*}
and
\begin{equation}
\bm{u}\equiv P_{n+1},\qquad\bm{u}_{k}\equiv P_{k},\qquad k=1,\ldots ,n.
\label{C7}
\end{equation}
The transformation between the jet coordinates on the respective leaves: $%
\mathcal{M}_{n,c}$ for $m=0$ and $\bar{\mathcal{M}}_{n,\bar{c}}$
for $m=1$, of the representations \eqref{L17a} and \eqref{L18a} of
$n$-th stationary KdV system and Vi\`{e}te's coordinates
$(\bm{q},\bm{p})$ is given, through the KdV co-symmetries
$\gamma_{i}$, as
\begin{equation}
q_{i}=\frac{1}{2}\gamma_{i},\qquad p_{i}=\frac{1}{2}\sum_{j=1}^{n}\bigl(%
G_{m}^{-1}\bigr)_{ij}(\gamma_{j})_{x},\qquad i=1,\ldots,n,\qquad (m=0,1),
\label{C8}
\end{equation}
where we make the identification $t_{1}\equiv x$. The transformation %
\eqref{C8} is a direct consequence of comparison of \eqref{uq} with %
\eqref{pn} through \eqref{C7} and the first Hamiltonian flow \eqref{V3} on $%
\bm{q}$: $\dot{\bm{q}}=G_{m}\bm{p}$. Notice that $q_{1}=\frac{1}{2}u$.
Obviously, one can obtain the transformation between jet coordinates and
separation variables from \eqref{C8} using the change of coordinates
\eqref{V1}.

Summing up the obtained results, we get the first part of Theorem~\ref{thm1},
the remaining part is included in Lemma~\ref{thm2}.

\begin{example}
\label{e1} The St\"{a}ckel representation of the first $KdV$ stationary
system on $\mathcal{M}_{2,c}$ (for $n=2$) is generated by the spectral curve
\begin{equation*}
\lambda^{5}+c\lambda^{2}+H_{1}\lambda+H_{2}=\mu^{2}.
\end{equation*}

The Hamiltonians \eqref{C4} in Vi\`{e}te's coordinates $(\bm{q},\bm{p})$ are
\begin{equation}  \label{ph}
\begin{split}
H_{1} & =\frac{1}{2}\bm{p}^{T}G_{0}\bm{p}+\mathcal{V}_{1}^{(5)}+c\mathcal{V}%
_{1}^{(2)}=2p_{1}p_{2}+q_{1}p_{2}^{2}-q_{1}^{4}-q_{2}^{2}+3q_{1}^{2}q_{2}+cq_{1},
\\
H_{2} & =\frac{1}{2}\bm{p}^{T}K_{2}G_{0}\bm{p}+\mathcal{V}_{1}^{(5)}+c%
\mathcal{V}%
_{1}^{(2)}=p_{1}^{2}+(q_{1}^{2}-q_{2})p_{2}^{2}+2q_{1}p_{1}p_{2}-q_{1}^{3}q_{2}+2q_{1}q_{2}^{2}+cq_{2},
\end{split}%
\end{equation}
where
\begin{equation*}
G_{0}=2%
\begin{pmatrix}
0 & 1 \\
1 & q_{1}%
\end{pmatrix}
,\qquad K_{1}=%
\begin{pmatrix}
1 & 0 \\
0 & 1%
\end{pmatrix}
,\qquad K_{2}=%
\begin{pmatrix}
0 & 1 \\
-q_{2} & q_{1}%
\end{pmatrix}
.
\end{equation*}

The related Lax operator \eqref{T2} has the form
\begin{equation}
\mathbb{L}=%
\begin{pmatrix}
-p_{2}\lambda-p_{1}-q_{1}p_{2} & \lambda^{2}+q_{1}\lambda+q_{2} \\
\lambda^{3}-q_{1}\lambda^{2}+(q_{1}^{2}-q_{2})%
\lambda-p_{2}^{2}-q_{1}^{3}+2q_{1}q_{2}+c & p_{2}\lambda+p_{1}+q_{1}p_{2}%
\end{pmatrix}
,  \label{p2a}
\end{equation}
and the generating matrices \eqref{T7} are
\begin{equation}
\mathbb{U}_{1}=%
\begin{pmatrix}
0 & 1 \\
\lambda-2q_{1} & 0%
\end{pmatrix}
,\qquad\mathbb{U}_{2}=%
\begin{pmatrix}
-p_{2} & \lambda+q_{1} \\
\lambda^{2}-q_{1}\lambda+q_{1}^{2}-2q_{2} & p_{2}%
\end{pmatrix}
.  \label{p2b}
\end{equation}

The respective Hamiltonian flows can be now obtained directly from the
Hamiltonian equations \eqref{V3} or equivalently Lax equations \eqref{T1}.
Thus, the first Hamiltonian flow has the form
\begin{equation}
\begin{pmatrix}
q_{1} \\
q_{2} \\
p_{1} \\
p_{2}%
\end{pmatrix}
_{t_{1}}=%
\begin{pmatrix}
2p_{2} \\
2p_{1}+2q_{1}p_{2} \\
-p_{2}^{2}+4q_{1}^{3}-6q_{1}q_{2}-c \\
2q_{2}-3q_{1}^{2}%
\end{pmatrix}
\label{p1}
\end{equation}
and the second one is given by
\begin{equation}
\begin{pmatrix}
q_{1} \\
q_{2} \\
p_{1} \\
p_{2}%
\end{pmatrix}
_{t_{2}}=%
\begin{pmatrix}
2p_{1}+2q_{1}p_{2} \\
2(q_{1}^{2}-q_{2})p_{2}+2q_{1}p_{1} \\
-2q_{1}p_{2}^{2}-2p_{1}p_{2}+3q_{1}^{2}q_{2}-2q_{2}^{2} \\
p_{2}^{2}+q_{1}^{3}-4q_{1}q_{2}-c%
\end{pmatrix}
.  \label{p2}
\end{equation}

The transformation to the jet coordinates is given by \eqref{C8}, thus
\begin{equation}
\begin{split}
& q_{1}=\frac{1}{2}\gamma_{1}\equiv\frac{1}{2}u,\qquad q_{2}=\frac{1}{2}%
\gamma_{2}\equiv\frac{1}{8}u_{xx}+\frac{3}{8}u^{2}, \\
& p_{1}=\frac{1}{4}(\gamma_{2})_{x}-\frac{1}{4}u(\gamma_{1})_{x}\equiv \frac{%
1}{16}u_{3x}+\frac{1}{4}uu_{x},\qquad p_{2}=\frac{1}{4}(\gamma_{1})_{x}\equiv%
\frac{1}{4}u_{x}.
\end{split}
\label{p3}
\end{equation}
Substituting \eqref{p3} to the first \eqref{p1} and the second flow
\eqref{p2} we obtain the equalities
\begin{equation*}
u_{t_{1}}=(\gamma_{1})_{x},\qquad u_{t_{2}}=(\gamma_{2})_{x},\qquad\gamma
_{3}+c=0,
\end{equation*}
which constitute the first representation of the $2$-th stationary KdV
system \eqref{E1}. Substituting \eqref{p3} to \eqref{p2a} and \eqref{p2b}
one reconstructs the respective Lax matrices \eqref{L7a} and \eqref{E2}.
Substituting \eqref{p3} to \eqref{ph} one reconstructs the integrals of
motion \eqref{E2h}.
\end{example}

\begin{example}
\label{e2} The St\"{a}ckel representation of the second $KdV$ stationary
system on $\bar{\mathcal{M}}_{2,\bar{c}}$ (for $n=2$) is generated by the
spectral curve
\begin{equation*}
\lambda^{4}+\bar{c}\lambda^{-1}+\bar{H}_{1}\lambda+\bar{H}_{2}=\lambda\mu^{2}.
\end{equation*}
Then, in Vi\`{e}te's coordinates $(\bm{q},\bm{p})$ we find the following
Hamiltonians
\begin{equation}
\begin{split}
\bar{H}_{1} & =\frac{1}{2}\bm{p}^{T}G_{1}\bm{p}+\mathcal{V}_{1}^{(4)}+\bar{c}
\mathcal{V}_{1}^{(-1)}=p_{1}^{2}-q_{2}p_{2}^{2}+q_{1}^{3}-2q_{1}q_{2}+\frac{\bar{c}}{q_{2}}, \\
\bar{H}_{2} & =\frac{1}{2}\bm{p}^{T}K_{2}G_{1}\bm{p}+\mathcal{V}_{2}^{(4)}+\bar{c}\mathcal{V}%
_{2}^{(-1)}=-2q_{2}p_{1}p_{2}-q_{1}q_{2}p_{2}^{2}+q_{1}^{2}q_{2}-q_{2}^{2}+%
\frac{\bar{c}q_{1}}{q_{2}},
\end{split}
\label{pph}
\end{equation}
where
\begin{equation*}
G_{1}=2%
\begin{pmatrix}
1 & 0 \\
0 & -q_{2}%
\end{pmatrix}%
\end{equation*}
and the Killing tensors are the same as in the previous example.

The related Lax operator and the generating matrices are given by
\begin{equation}
\mathbb{L}=%
\begin{pmatrix}
-p_{1}\lambda+q_{2}p_{2} & \lambda^{2}+q_{1}\lambda+q_{2} \\
\lambda^{3}-q_{1}\lambda^{2}+(q_{1}^{2}-q_{2})\lambda+(\bar{c}%
-q_{2}^{2}p_{2}^{2})q_{2}^{-1} & p_{1}\lambda-q_{2}p_{2}%
\end{pmatrix}
\label{pp2a}
\end{equation}
and
\begin{equation}
\mathbb{U}_{1}=%
\begin{pmatrix}
0 & 1 \\
\lambda-2q_{1} & 0%
\end{pmatrix}
,\qquad\mathbb{U}_{2}=%
\begin{pmatrix}
-p_{1} & \lambda+q_{1} \\
\lambda^{2}-q_{1}\lambda+q_{1}^{2}-2q_{2} & p_{1}%
\end{pmatrix}
.  \label{pp2b}
\end{equation}

The respective Hamiltonian flows are
\begin{equation}
\begin{pmatrix}
q_{1} \\
q_{2} \\
p_{1} \\
p_{2}%
\end{pmatrix}
_{t_{1}}=%
\begin{pmatrix}
2p_{1} \\
-2q_{2}p_{2} \\
-3q_{1}^{2}+2q_{2} \\
p_{2}^{2}+2q_{1}+\bar{c}q_{2}^{-2}%
\end{pmatrix}
\label{pp1}
\end{equation}
and
\begin{equation}
\begin{pmatrix}
q_{1} \\
q_{2} \\
p_{1} \\
p_{2}%
\end{pmatrix}
_{t_{2}}=%
\begin{pmatrix}
-2q_{2}p_{2} \\
-2q_{1}q_{2}p_{2}-2q_{2}p_{1} \\
q_{2}p_{2}^{2}-2q_{1}q_{2}-\bar{c}q_{2}^{-1} \\
2p_{1}p_{2}+q_{1}p_{2}^{2}-q_{1}^{2}+2q_{2}+\bar{c}q_{1}q_{2}^{-2}%
\end{pmatrix}
.  \label{pp2}
\end{equation}

The transformation to the jet coordinates \eqref{C8} takes the form
\begin{equation}
\begin{split}
& q_{1}=\frac{1}{2}\gamma_{1}\equiv\frac{1}{2}u,\qquad q_{2}=\frac{1}{2}%
\gamma_{2}\equiv\frac{1}{8}u_{xx}+\frac{3}{8}u^{2}, \\
& p_{1}=\frac{1}{4}(\gamma_{1})_{x}\equiv\frac{1}{4}u_{x},\qquad p_{2}=-%
\frac{1}{2}\frac{(\gamma_{2})_{x}}{\gamma_{2}}\equiv-\frac{1}{2}\frac {%
u_{3x}+6u u_{x}}{u_{xx}+3u^{2}}.
\end{split}
\label{pp3}
\end{equation}
Substituting \eqref{pp3} to the first \eqref{pp1} and the second flow
\eqref{pp2} we obtain the equalities
\begin{equation*}
u_{t_{1}}=(\gamma_{1})_{x},\qquad u_{t_{2}}=(\gamma_{2})_{x},\qquad\frac{1}{2%
}\gamma_{2}(\gamma_{2})_{xx}-\frac{1}{4}(\gamma_{2})_{x}^{2}+u%
\gamma_{2}^{2}+4\bar{c}=0,
\end{equation*}
which constitute the second representation of the $2$-th stationary KdV
system \eqref{E3}. Substituting \eqref{pp3} to \eqref{pp2a}, \eqref{pp2b}
and \eqref{pph} one reconstructs the respective Lax matrices \eqref{L7a},
\eqref{E4} and the integrals of motion \eqref{E4h}.
\end{example}

\section{Miura map and bi-Hamiltonian representations of stationary systems}

We know that the St\"{a}ckel systems associated with the curves \eqref{C1}
and \eqref{C3} are two different representations of the same $n$-th stationary KdV
system on leaves of two mutually transversal Hamiltonian foliations of the stationary manifold
$\mathcal{M}_{n}$. These representations are non-equivalent unless we extend the considered
phase space of each of these systems to the $(2n+1)$-dimensional phase space
$\mathcal{M}_{n}\cong M\oplus\mathbb{R}$, where $M$ now plays the role of the respective
leaf for both foliations. Consider thus the curve \eqref{C1}
\begin{equation}
\lambda^{2n+1}+\sum_{k=0}^{n}H_{k}\lambda^{n-k}=\mu^{2},  \label{D1}
\end{equation}
on the phase space  $\mathcal{M}_{n}$, parameterized by the extended Vi\`{e}te's coordinates $(\bm{q},\bm{p},c)$. It leads to Hamiltonians~\eqref{C2}, $H_{k}=H_{k}(\bm{q},\bm{p},c)$,
explicitly depending on the additional trivial integral of motion,
$H_{0}:=c$.

Consider also the curve \eqref{C3}
\begin{equation}
\bar{\lambda}^{2n}+\sum_{k=1}^{n+1}\bar{H}_{k}\bar{\lambda}^{n-k}=\bar
{\lambda}\bar{\mu}^{2},  \label{D2}
\end{equation}
yielding the Hamiltonians \eqref{D2}, $\bar{H}_{k}=\bar{H}_{k}(\bm{Q},\bm{P},
\bar{c})$, explicitly depending on the additional trivial integral of motion,
$\bar{H}_{n+1}:=\bar{c}$. These Hamiltonians are defined on the same extended phase space $\mathcal{M}
_{n}$, parameterized this time by the extended Vi\`{e}te's coordinates
$(\bm{Q},\bm{P}, \bar{c})$. Having established this notation, we can now formulate
the following lemma.

\begin{lemma}
\label{thm2} The Hamiltonian equations \eqref{V3} associated with the St\"{a}ckel systems defined by the curves \eqref{D1} and \eqref{D2} are related by
the Miura map on $\mathcal{M}_{n}$
\begin{align}
q_{i} & =Q_{i},\qquad i=1,...,n  \notag \\
p_{1} & =-\bigl(Q_{1}P_{1}+Q_{2}P_{2}+\ldots+Q_{n}P_{n}\bigr),\qquad
p_{i}=P_{i-1},\qquad i=2,...,n,  \label{M1} \\
c & =\bar{H}_{1}(\bm{Q},\bm{P},\bar{c})  \notag
\end{align}
and its inverse
\begin{align}
Q_{i} & =q_{i},\qquad i=1,...,n  \notag \\
P_{i} & =p_{i+1},\qquad i=1,...,n-1,\qquad P_{n}=-\frac{1}{q_{n}}\bigl (
p_{1}+q_{1}p_{2}+\ldots+q_{n-1}p_{n}\bigr),  \label{M2} \\
\bar{c} & =H_{n}(\bm{q},\bm{p},c).  \notag
\end{align}
\end{lemma}

\begin{proof}
It is simple to observe that the curves \eqref{D1} amd \eqref{D2} are
related by the relations
\begin{equation}
\lambda=\bar{\lambda},\qquad\mu=\bar{\lambda}\bar{\mu}  \label{D3}
\end{equation}
and the following identification between Hamiltonians:
\begin{equation}  \label{D4}
\begin{split}
c\equiv H_{0}(\bm{q},\bm{p},c) & =\bar{H}_{1}(\bm{Q},\bm{P},\bar{c}), \\
H_{i}(\bm{q},\bm{p},c) &
=\bar{H}_{i+1}(\bm{Q},\bm{P},\bar{c}),\qquad
i=1,\ldots,n-1, \\
H_{n}(\bm{q},\bm{p},c) &
=\bar{H}_{n+1}(\bm{Q},\bm{P},\bar{c})\equiv \bar{c}.
\end{split}
\end{equation}
Let us notice that the map \eqref{D3} translates immediately to the transformation
between separation variables and so the Vi\`{e}te's ones too, where the
coordinates on the configuration space are preserved, hence $\bm{q}=\bm{Q}$.
Now, the connection between coordinates on the extended phase space is
consequence of \eqref{D4} and the equivalence $\dot{\bm{q}}=\dot{\bm{Q}}$.
Thus, from \eqref{V3} for $k=1$ we have
\begin{equation*}
\bm{p}=G_{0}^{-1}G_{1}\bm{P}\equiv R^{T}\bm{P}\qquad\Longleftrightarrow
\qquad\bm{P}=G_{1}^{-1}G_{0}\bm{P}\equiv(R^{-1})^{T}\bm{p},
\end{equation*}
where $R$, given by \eqref{V5}, is the matrix representation, valid only in
Vi\`{e}te's coordinates, of the special conformal Killing tensor $L$ \cite {blaszak2007}.
Alternatively, the Miura map can be constructed through the
respective changes of coordinates from the separation variables to
Vi\`{e}te's coordinates \eqref{V1} using the relations \eqref{D3}.
\end{proof}

The Miura map (\ref{M1}), or equivalently its inverse (\ref{M2}), represents
the non-canonical transformation between two sets $(\bm{q},\bm{p},c)$ and $(
\bm{Q},\bm{P},\bar{c})$ of canonical coordinates on the extended
phase space $\mathcal{M}_{n}$. In consequence both St\"{a}ckel
representations are bi-Hamiltonian on $\mathcal{M}_{n}$
\cite{blaszak2009} and the Miura map
transforms the canonical Poisson structure in parametrization $(\bm{Q},\bm{P}
,\bar{c})$ of one St\"{a}ckel system into non-canonical Poisson
structure in parametrization $(\bm{q},\bm{p},c)$ of the second
one.

\begin{example}
\label{e3} Consider again the case $N=n=2$. The Miura map \eqref{M1} on the stationary manifold
$\mathcal{M}_{2}$ has the form:
\begin{equation}\label{mp41}
\begin{split}
q_{1}& =Q_{1},\qquad q_{2}=Q_{2},\qquad
p_{1}=-Q_{1}P_{1}-Q_{2}P_{2},\qquad p_{2}=P_{1},  \\
c& =\bar{H}_{1}(\bm{Q},\bm{P},\bar{c})
\equiv P_{1}^{2}-Q_{2}P_{2}^{2}+Q_{1}^{3}-2Q_{1}Q_{2}+\frac{\bar{c}}{Q_{2}},
\end{split}
\end{equation}
where $\bar{H}_{1}$ is given by (\ref{pph}). This yields the bi-Hamiltonian
representation of St\"{a}ckel systems, from
Example~\ref{e1}, on the extended phase space parametrized by
$(q_{1},q_{2},p_{1},p_{2},c)$:
\begin{equation*}
\pi _{0}dc=0,\qquad \pi _{0}dH_{1}=\pi _{1}dc,\qquad \pi _{0}dH_{2}=\pi
_{1}dH_{1},\qquad 0=\pi _{1}dH_{2},
\end{equation*}
where $\pi _{0}$ is a canonical Poisson tensor, with matrix representation:
\begin{equation*}
 \pi _{0} =
\begin{pmatrix}
0 & 0 & 1 & 0 & 0 \\
0 & 0 & 0 & 1 & 0 \\
-1 & 0 & 0 & 0 & 0 \\
0 & -1 & 0 & 0 & 0 \\
0 & 0 & 0 & 0 & 0%
\end{pmatrix},
\end{equation*}
and $\pi _{1}$ is a non-canonical one, generated by the Miura map (\ref{mp41}),
with matrix representation:
\begin{equation*}
  \pi _{1}=\begin{pmatrix}
 0 & 0 & -q_1 & 1 & 2 p_2 \\
 0 & 0 & -q_2 & 0 & 2p_1+2q_1p_2 \\
 q_1 & q_2 & 0 & -p_2 & -p_{2}^{2}+4q_{1}^{3}-6q_{1}q_{2}-c \\
 -1 & 0 & p_2 & 0 & 2 q_2-3 q_1^2 \\
 -2 p_2 & -2p_1-2q_1p_2 & p_{2}^{2}-4q_{1}^{3}+6q_{1}q_{2}+c & -2 q_2+3 q_1^2 & 0
\end{pmatrix},
\end{equation*}
where in the last column naturally appear components of the Hamiltonian vector field \eqref{p1}.

The bi-Hamiltonian representation of St\"{a}ckel systems from Example \ref{e2}
can be constructed in a similar fashion.
\end{example}

\section{Conclusions and further research}

In this article we have systematized the already existing knowledge about stationary KdV systems, but also
added new facts: we have shown explicit maps between the jet variables of each stationary KdV system
and separation coordinates (via Viete's coordinates) of its two St\"{a}ckel representations on $2n$-dimensional phase space. These two non-equivalent St\"{a}ckel representations of the same stationary KdV system, when considered on the extended $(2n+1)$-dimensional phase space $\mathcal{M}_n$, are connected by a Miura map, which leads to
bi-Hamiltonian formulation of the stationary KdV system, known from the literature.
Thus, we have proved that different St\"{a}ckel representations of stationary KdV systems are related with different foliations of stationary manifolds $\mathcal{M}_n$.

Based on results of this article, we will attempt to develop a
similar theory for the coupled ($N$-field) KdV hierarchy. This
will be a topic of an upcoming article.

\end{document}